\documentclass[aps,prb,twocolumn,groupedaddress,amsmath]{revtex4}

\usepackage{dcolumn,graphicx}

\newcommand\acis{Adv. Colloid Interface Sci.}
\newcommand\acp{Adv. Chem. Phys.}
\newcommand\cp{Chem. Phys.}
\newcommand\cpr{Comp. Phys. Rep.}
\newcommand\jasa{J. Am. Stat. Assoc.}
\newcommand\jpa{J. Phys. A}
\newcommand\jpcm{J. Phys. Cond. Mat.}
\newcommand\jsp{J. Stat. Phys.}
\newcommand\molp{Mol. Phys.}
\newcommand\sci{Science}

\begin{document}
\title{Phase diagram of the adhesive hard sphere fluid}
\author{Mark A.~Miller}
\affiliation{University Chemical Laboratory, Lensfield Road, Cambridge CB2 1EW, U.~K.}
\author{Daan Frenkel}
\affiliation{FOM Institute for Atomic and Molecular Physics, Kruislaan 407,
1098 SJ Amsterdam, The Netherlands}
\date{\today}

\begin{abstract}
The phase behavior of the Baxter adhesive hard sphere fluid has been
determined using specialized Monte Carlo simulations.  We give a
detailed account of the techniques used and present data for the
fluid--fluid coexistence curve as well as parametrized fits
for the supercritical equation of state
and the percolation threshold.  These properties are compared
with the existing results of Percus--Yevick theory for this system.
\end{abstract}

\maketitle

\section{Introduction}

For a pure substance to be capable of exhibiting equilibrium
between two fluid phases---a gas and a liquid---the interaction
between its particles must include an attractive component.
Possibly the simplest model to incorporate attraction is the
square-well potential, where particles present each other with a
hard core and a uniform region of favorable potential energy up to
an abrupt distance threshold beyond which there is no interaction.
The square-well model has two intrinsic length scales: the
hard-core diameter and the width of the attractive well. The ratio
of these lengths determines the phase behavior.
\par
In 1968, Baxter introduced a model fluid containing attraction but
having only one length scale, the hard-core diameter.\cite{Baxter68a}
In this model, the attraction takes the form of an adhesive interaction when
particles are actually in contact, and therefore has an
effectively infinitesimal range. Baxter produced an analytic
equation of state for this ``adhesive hard sphere'' model using
the Percus--Yevick framework and compressibility equation, showing
that the model possesses a critical point below which gas--liquid
equilibrium may be observed.  A second Percus--Yevick equation of
state---derived from the energy equation---soon
followed,\cite{Watts71a} with a strikingly different critical
point from the compressibility-equation prediction.
\par
The Percus--Yevick framework also furnishes the cluster size distribution.\cite{Chiew83a}
The surface adhesion provides an unambiguous definition of clusters as
connected sets of mutually touching particles.  At sufficiently low density,
the fluid consists only of monomers and finite clusters, but above a
temperature-dependent density threshold, clusters tend to aggregate
into amorphous macroscopic structures.  The system then contains sample-spanning
infinite clusters and is said to percolate.  The percolation threshold, at which the
divergence of the cluster size occurs, has been located within the Percus--Yevick
approximation by Chiew and Glandt.\cite{Chiew83a}
\par
Baxter regarded the adhesive hard sphere model as a highly idealized
manifestation of a general potential with a repulsive core and an attractive
tail.  However, it has since been suggested\cite{Menon91a} that it is quite
a reasonable
description of colloidal systems, where the particles interact on length
scales much smaller than their own size.  The gas and liquid are
then more appropriately regarded as low- and high-density colloidal fluid phases.
Percolation can be an observable phenomenon in colloidal systems,
and may sometimes be detected directly, such as by a sudden increase
in electrical conductivity across the sample.\cite{Chen94a}
The Baxter model is particularly appealing in the light of recent
theoretical\cite{Dawson01a} and experimental\cite{Pham02a}
work that predicts and confirms the existence of a re-entrant
glass--liquid--glass transition in colloidal systems with short-range
attractive forces.  For very short-range attraction there is also a
glass--glass transition between a caged ``repulsive'' glass and an
energetic ``attractive'' glass.
\par
A major obstacle in the application of the adhesive hard sphere model
to the description of experimental data is that the model's phase
behavior is only known through the approximate and conflicting theoretical
results of Percus--Yevick theory.  Computer
simulations\cite{Seaton87b,Seaton87a,Kranendonk88a,Lee01c}
have provided some information on percolation, but have not
tackled the issue of phase coexistence.
The purpose of the present contribution is to provide a better knowledge
of the adhesive hard sphere phase diagram using computer simulation.
We describe in detail the
special Monte Carlo techniques employed (Section
\ref{techniques}) and supply numerical results for the
fluid--fluid coexistence curve, the percolation threshold, and the
supercritical equation of state (Section \ref{results}).  We begin,
in Section \ref{model}, with a definition of the model itself.
It will be seen that this definition involves taking a limit that
introduces a singularity into the potential energy function---an
alarming prospect at first sight.  The model in its strictest
interpretation is indeed pathological,\cite{Stell91a}
but the problem is rather more subtle than it first appears, and in fact
starts to set in before the full limit is taken.
A proper discussion of this point is deferred to Section \ref{pathology}, where the
difficulties can be put into the context of the simulation methodology.

\section{The Adhesive Hard Sphere Model\label{model}}

The adhesive hard sphere model is defined by the pair potential function\cite{Baxter68a}
\begin{equation}
U(r)/kT =
   \begin{cases}
   \infty & 0 \le r < \sigma\\
   \ln [12\tau(1-\sigma/d)] & \sigma \le r < d\\
   0 & d \le r,
   \end{cases}
\label{potential}
\end{equation}
in the limit $d\to\sigma$, where $kT$ is the thermal energy and
$r$ the particle separation.  Eq.~(\ref{potential}) describes a
square well with hard-core diameter $\sigma$, width $d-\sigma$,
and depth controlled by the parameter $\tau$.

Equation (\ref{potential}) couples the depth of the well
to its width in such a
way that the total Boltzmann weight of bound configurations remains
finite as the well width is reduced to zero:
\begin{align}
\lim_{d\to\sigma}\int_\sigma^d e^{-U(r)/kT}4\pi r^2 dr
&=\lim_{d\to\sigma}\frac{d}{12\tau}\frac{4\pi}{3}(d^2+\sigma d+\sigma^2)
\nonumber\\
&=\frac{\sigma}{12\tau}4\pi\sigma^2
\end{align}
Hence, in the limit $d\to\sigma$, which corresponds to an infinitesimally
narrow but infinitely deep well, one can write
\begin{equation}
e^{-U(r)/kT} = \Theta(r-\sigma) + \frac{\sigma}{12\tau}\delta(r-\sigma).
\label{boltzmann}
\end{equation}
The Heaviside step function, $\Theta$, accounts for the hard-core
repulsion, while the Dirac $\delta$ represents the surface
adhesion.  Eq.~(\ref{boltzmann}) is our working definition of the
Baxter model.

In the limit $\tau\to\infty$, Eq.~(\ref{boltzmann}) reduces to the hard
sphere potential, so $\tau^{-1}$ measures the strength of adhesion.
Alternatively, $\tau$ itself can be regarded as the effective temperature,
since the true thermal energy $kT$ has been absorbed into the definition
of the potential and does not appear explicitly in the right-hand side
of Eq.~(3).  In the following, we shall refer to $\tau$ as the temperature.
\par
The configuration space of an $N$-particle adhesive hard sphere system can be
notionally divided into subspaces within each of which the number of
binary contacts is constant.  Let
\begin{equation}
\Omega_{NM}=\frac{1}{N!}
\left(\frac{\sigma}{12}\right)^M\int\delta_M({\bf r}^{3N})
\prod_{i<j}^N \Theta(r_{ij}-\sigma)
 d{\bf r}^{3N},
\label{dos}
\end{equation}
where $r_{ij}$ is the separation of particles $i$ and $j$.  The
integral extends over all particle coordinates, but the only
configurations that make a non-zero contribution are those in
which no particles overlap and the total number of binary contacts
is $M$.  The latter requirement is enforced by the function
$\delta_M({\bf r}^N)$.
We point out that the presence of delta-functions in the Boltzmann
factor of a pair of adhesive hard spheres should not be
interpreted as a holonomic constraint on the Lagrangian of the
system but rather as the limit of an increasingly stiff bond. In
fact, if the bonds between adhesive hard spheres acted as
holonomic constraints, then the integrand in Eq.~(\ref{dos}) would
include a Jacobian as a result of the integration over the
(constrained) momenta (see e.g.~Ref.~\onlinecite{Ciccotti86a}), not to
mention the fact that the dimensionality of the phase space would
in that case depend on the number of bonds.
Although we will not need to manipulate an explicit form for
$\delta_M({\bf r}^N)$, it may be formally expressed as a sum of
$M$-fold products of $\delta$ functions:
\begin{equation*}
\delta_M({\bf r}^{3N})=\sum^N_{\{i_1,\dots i_{2M}\}}
\delta(r_{i_1,i_2}-\sigma)\dots
\delta(r_{i_{2M-1},i_{2M}}-\sigma),
\end{equation*}
where $r_{ij}$ is the separation of particles $i$ and $j$. The sum is over
all combinations of the $2M$ indices that give distinct products of
$\delta$ functions but neither generate constraints
on a `self-separation' (such as $r_{ii}$) nor duplicate a constraint on
any given separation within a single term in the sum.
\par
The configuration integral
now takes the succinct form of Eq.~(\ref{qn}), which emphasizes the
role of $\Omega_{NM}$ as an effective density of states:
\begin{align}
Z_N(\tau)&= \frac{1}{N!}\int\exp\left[-\sum^N_{i<j}U(r_{ij})/kT\right]d{\bf r}^{3N}\nonumber\\
         &= \sum_M\Omega_{NM}\tau^{-M}.
\label{qn}
\end{align}
This notation will be useful when we consider histogram reweighting.

\section{Simulation\label{techniques}}

\subsection{Canonical Monte Carlo}

The singularity in Eq.~(\ref{boltzmann}) means that the conventional Monte
Carlo algorithm, in which the test particle undergoes random displacements,
is not applicable.  The probability of a trial move bringing two particles
exactly into contact vanishes, and the energy change associated with breaking
such a contact is formally infinite.  Nevertheless, the integrals in
Eq.~(\ref{dos}) are finite (though see Section \ref{pathology})
and there is an equilibrium between
states with different total numbers of contacts, even if the time scales
for establishing the equilibrium diverge as the limit of vanishing potential
range is taken.\cite{Stell91a}
In simulations of the adhesive hard sphere model it is therefore necessary
to employ moves that explicitly make or break contacts.
\par
For our canonical simulations we adopt the method pioneered by Seaton and
Glandt,\cite{Seaton86a,Seaton87a} and refined by Kranendonk and
Frenkel.\cite{Kranendonk88a}  In a conventional Monte Carlo simulation
the Boltzmann weights of configurations are compared,
where a configuration means a specification of all particle coordinates.
In contrast, simulations of the adhesive hard sphere model proceed by
comparing the weights of different coordination states.  Here we use the phrase
coordination state to mean a specification of which particles are touching
each other.  In general, many configurations are compatible with a given
coordination state.  Since the Hamiltonian depends only on the number of
contacts in the system, all configurations belonging to the same
coordination state have the same statistical weight.
With these concepts in mind, the key features of the algorithm
may be outlined as follows.
\begin{enumerate}
\item A particle (the ``test particle'') is chosen at random.
\item A list is made of the possible coordination states of the test particle
with other particles in the system, keeping the positions of the latter fixed.
The list always contains one item
in which the test particle makes no contacts at all, and $N-1$ items
where it touches just one of the other particles in the system.  It also
enumerates pairs of particles that are sufficiently close for the test
particle to touch both simultaneously, as well as triplets
compatible with three simultaneous contacts to the test particle.
\item For each coordination state in the list, a transition probability is
calculated.\label{calcstep}
\item A coordination state is chosen at random with weight in proportion to
its transition probability.\label{trstep}
\item A uniformly distributed random configuration belonging to the
chosen coordination state is chosen.\label{example}
\item If the configuration generates a hard-sphere overlap with another
particle in the system, the move is rejected, otherwise it is accepted.\label{accstep}
\end{enumerate}
We now show how to calculate the transition probabilities in step \ref{calcstep}
so that repeated
application of this procedure generates the correct Boltzmann distribution
of configurations.  The total relative unnormalized weight of a coordination
state of the test particle is given by the integral of the Boltzmann factor over
all configurations belonging to that coordination state.  For the coordination state
involving no contacts with other particles, this weight is the free volume accessible
to the test particle, which we label $t$:
\begin{equation}
W^{(0)}=\int_V \prod_{i\ne t}^N\Theta(r_{ti}-\sigma)d{\bf r}_t,
\label{W0}
\end{equation}
$V$ being the volume of the simulation box.
For a coordination state involving contact of the test particle with one other particle,
$a$, the test particle is constrained to the accessible part of the surface of $a$:
\begin{equation}
W^{(1)}=\frac{\sigma}{12\tau}\int_V \delta(r_{ta}-\sigma)\prod_{i\ne t}^N\Theta(r_{ti}-\sigma)d{\bf r}_t.
\end{equation}
Similarly, for coordination states involving simultaneous contact with two particles
($a$ and $b$) or three particles ($a$, $b$ and $c$), we have
\begin{equation}
W^{(2)}=\left(\frac{\sigma}{12\tau}\right)^2
\int_V \delta(r_{ta}-\sigma)\delta({r_{tb}-\sigma})\prod_{i\ne t}^N\Theta(r_{ti}-\sigma)d{\bf r}_t
\end{equation}
and
\begin{multline}
W^{(3)}=\left(\frac{\sigma}{12\tau}\right)^3
\int_V \delta(r_{ta}-\sigma)\delta({r_{tb}-\sigma})\delta(r_{tc}-\sigma)\times\\
\prod_{i\ne t}^N\Theta(r_{ti}-\sigma)d{\bf r}_t.
\label{W3}
\end{multline}
Moves that create or destroy more than three contacts simultaneously are not
considered in the algorithm.  A particle can, however, attain a coordination
number greater than three through suitable combinations of moves where three
or fewer contacts are created.
\par
If Eqs.~(\ref{W0}) to (\ref{W3}) could be evaluated conveniently, the canonical distribution
could be generated directly by choosing a coordination state at random using these weights.
The transition probability would be
\begin{equation}
\alpha_{ij}=\frac{W_j}{\sum_{k=1}^n W_k},
\end{equation}
where the possible coordination states of the test particle have been numbered from 1
to $n$, and the weight $W_j$ of state $j$ has been evaluated using the appropriate
formula from Eqs.~(\ref{W0}) to (\ref{W3}).  Note that $\alpha_{ij}$ does not depend
on $i$, and that all moves would be accepted, since new configurations never generate
overlaps and are already chosen with the desired probability distribution.
\par
The product of step functions that appears in Eqs.~(\ref{W0}) to (\ref{W3}) to
preclude overlaps with the remaining particles in the system, makes evaluation
of the weights very difficult.  Following Seaton and Glandt,\cite{Seaton86a,Seaton87a}
we consider instead the modified weights, in which the overlap is initially ignored.
The weight of the coordination state where the test particle makes no contacts is
now a free integral over the container volume:
\begin{equation}
W'^{(0)}=\int_V d{\bf r}_t.
\end{equation}
The weight for a one-contact coordination state is an integral over the complete
sphere of radius $\sigma$ surrounding a particle:
\begin{equation}
W'^{(1)}=\frac{\sigma}{12\tau}\int_V \delta(r_{ta}-\sigma) d{\bf r}_t.
\label{Wp1}
\end{equation}
For two simultaneous contacts we integrate around a circle of configurations:
\begin{equation}
W'^{(2)}=\left(\frac{\sigma}{12\tau}\right)^2
\int_V \delta(r_{ta}-\sigma)\delta({r_{tb}-\sigma}) d{\bf r}_t
\end{equation}
and for three contacts, only two configurations (located symmetrically above
and below the plane defined by $a$, $b$ and $c$) contribute:
\begin{equation}
W'^{(3)}=\left(\frac{\sigma}{12\tau}\right)^3
\int_V \delta(r_{ta}-\sigma)\delta({r_{tb}-\sigma})\delta(r_{tc}-\sigma) d{\bf r}_t.
\label{Wp3}
\end{equation}
The integrals in Eqs.~(\ref{Wp1}) to (\ref{Wp3}) can be evaluated
by suitable series of variable changes.  Alternatively, one may
consider first a square-well potential of finite width.
Coordination states with 1, 2 and 3 contacts then restrict the
test particle's center-of-mass to, respectively, a spherical
shell, a torus with diamond-shaped cross-section, and a pair of
parallelepipeds.  Taking the limit of infinitesimal well width
then produces the same results as direct integration of the delta
functions.  One obtains
\begin{align}
\label{Wp0e}
W'^{(0)}&=V,\\
W'^{(1)}&=\frac{\pi\sigma^3}{3\tau},\\
W'^{(2)}&=\frac{\pi\sigma^4}{72\tau^2 r_{ab}},\\
\label{Wp3e}
W'^{(3)}&=\left(\frac{\sigma}{12\tau}\right)^3
   \frac{2r_{ta} r_{tb} r_{tc}}{|({\bf r}_{ta}\times{\bf r}_{tb}).{\bf r}_{tc}|}.
\end{align}
Unlike previous workers,\cite{Seaton87a,Lee93a} we consider moves to all triplets
of particles capable of simultaneously touching the test particle, not just to those
triplets where $a$, $b$ and $c$ are all mutually touching.  The coordination states
neglected by the latter approach can be reached by combinations of other moves, but
including the direct moves should make the algorithm more efficient at
low temperature or high density, where the average coordination number is expected to
be high.
\par
It is the weights in Eqs.~(\ref{Wp0e}) to (\ref{Wp3e}) that we use for the transition
probability in step \ref{trstep} of the Monte Carlo algorithm:
\begin{equation}
\alpha'_{ij}=\frac{W'_j}{\sum_{k=1}^n W'_k}.
\label{modalpha}
\end{equation}
To show that this choice satisfies detailed balance and leads to the correct Boltzmann
distribution, let $f_j=W_j/W'_j$ be the fraction of configurations belonging to
coordination state $j$ that are not blocked by overlap with other particles.  The
rejection of overlaps in step \ref{accstep} of the algorithm means that a trial move
to coordination state $j$ will be accepted with probability $p^{\rm acc}_{ij}=f_j$.
The overall probability of moving to state $j$ from state $i$ is therefore
\begin{equation}
\pi_{ij}=\alpha'_{ij}p^{\rm acc}_{ij}=\frac{W_j}{\sum_{k=1}^n W_k/f_k}.
\label{piij}
\end{equation}
Note that the sum in the denominator of Eq.~(\ref{piij}) depends on the positions of
all the particles except the test particle.
Since states $i$ and $j$ differ only by the position of the test particle,
this sum is the same for the reverse move, so that
$W_i\pi_{ij}=W_j\pi_{ji}$, and detailed balance is satisfied.
Recalling that we do not attempt moves that create more than three contacts simultaneously,
detailed balance also requires that an attempted displacement of any particle that has
accumulated more than three contacts is automatically rejected.
\par
Having chosen a coordination state in step \ref{trstep} of the algorithm, step
\ref{example} requires a uniformly distributed random configuration from this
state to be generated.  For coordination states involving 0, 1, 2 or 3 contacts
one must therefore select, respectively, a random point in the simulation box,
a random point on a sphere surrounding particle $a$, a random point on the circle
of contact with particles $a$ and $b$, or one of the two points lying a distance
$\sigma$ from each of $a$, $b$ and $c$.
\par
In practice, it is too time-consuming to enumerate all possible coordination states
of the test particle, and this task scales unfavorably with the system size.  It is
therefore desirable to restrict movement of the test particle to the vicinity of its
original position, so that only a subset of the possible coordination states need
be identified and considered.  However, as discussed by Seaton and Glandt,\cite{Seaton86a}
care must be taken since the normalizing sum in Eq.~(\ref{piij}) then depends on the
original position of the particle.  Kranendonk and Frenkel\cite{Kranendonk88a} have
prescribed a solution to the problem: a random point is chosen within
a sphere of radius $r_{\rm test}$ centered on the original position of the test particle,
and this point is used as the center of a second sphere (the test sphere), also of radius $r_{\rm test}$.
Displacements of the test particle are then considered to coordination states within
the test sphere.  The chance of choosing the same test sphere in the reverse
move is identical, so that the modified scheme maintains detailed balance.
\par
The test sphere introduces another complication in that it may intersect some of the
spherical and circular contact surfaces available to the test particle, thereby excluding
a fraction of the configurations belonging to the corresponding coordination states.
One must therefore calculate what fraction of each contact surface lies within
the test sphere, and modify the weights in Eq.~(\ref{modalpha}) by these fractions
when calculating the transition probabilities in step
\ref{calcstep} of the algorithm.  In step \ref{example}, one must then choose a
uniformly distributed random configuration restricted to the part of the surface that lies
within the test sphere.

\subsection{Grand Canonical Monte Carlo}

Since two phases can only coexist when their pressures are equal, computational
studies of coexistence often employ an ensemble where the size of the simulation cell
can fluctuate in response to the pressure.  The coordinates of the particles or
the cluster centers-of-mass are typically scaled in
proportion to the change in the cell length.
In the adhesive hard sphere fluid,
clusters can percolate even at moderately low density.  A percolating
cluster spans the simulation cell, so that increasing or decreasing the cell size in
the presence of such a cluster would always
involve breaking a contact or generating an overlap between at least two
particles.  Since overlaps are forbidden, and the breaking of a bond is energetically
prohibited, all trial changes in the cell size would be rejected.  Isobaric-isothermal
and Gibbs-ensemble simulations therefore fail as soon as the percolation threshold
is crossed.
\par
To study density fluctuations, we work instead in the grand canonical ensemble, where the
volume of the simulation is fixed, but the number of particles fluctuates in response to the
imposed chemical potential.  To avoid the formally infinite change in potential energy
associated with the creation and destruction of contacts between particles, we only consider
the insertion and removal of particles whose coordination number is zero.  The procedure is
therefore identical to that used for ordinary hard spheres,\cite{Adams74a} with the slight
modification that a removal must be rejected if the chosen particle happens to have a non-zero
coordination number.
In our grand canonical simulations, particle insertion and removal steps are attempted with
equal probability totaling 45\%.
\par
In an attempt to accelerate equilibration, we also perform cluster translation moves with
probability 5\%.  A particle is chosen at random, and then all particles in the cluster to
which it belongs are translated by the same amount.  The maximum translation in each Cartesian
direction is inversely proportional to the number of particles in the cluster.

\subsection{Parallel Tempering}

At low temperature and high density, the formation of large, highly-coordinated clusters
drastically slows down the equilibration of the simulations.  To help overcome this problem,
we employ the parallel tempering scheme of Geyer\cite{Geyer95a} in the grand canonical ensemble.
In our implementation, several runs are performed simultaneously at the same temperature and a
series of increasing chemical potentials.  At sufficiently low chemical potential, one can always recover
an ergodic dilute gas.  Parallel tempering involves periodic attempts to exchange the configurations
of pairs of adjacent runs in the hierarchy of chemical potentials.  The acceptance probability of
such moves is given below and ensures that the correct grand canonical distribution is obtained
for each run individually.  The advantage of swapping configurations is that a large cluster
formed at high chemical potential breaks down when transferred to lower chemical potential,
while a new cluster is built up from the configuration received by the run at higher chemical
potential.
\par
Using Eq.~(\ref{qn}), we can write the grand partition function of the adhesive hard sphere
model as
\begin{equation}
\Xi(\tau,z)=\sum_{N=0}^\infty\left(\frac{e^{\mu/kT}}{\Lambda^3}\right)^N Z_N(\tau)
=\sum_{N,M} \Omega_{NM} z^N \tau^{-M},
\end{equation}
where the chemical potential $\mu$ enters through the
activity $z=\Lambda^{-3}\exp(\mu/kT)$ and $\Lambda$ is the
thermal de Broglie wavelength.
The probability of observing the system
in a configuration with $N$ particles and $M$ contacts under conditions of
effective temperature $\tau$ and activity $z$ is
\begin{equation}
P_{NM}(\tau,z)=\Omega_{NM} z^N \tau^{-M} / \Xi(\tau,z).
\label{pnm}
\end{equation}
The joint probability of a system having $N_1$ particles and $M_1$ contacts
at temperature $\tau$ and activity $z_1$ at the same time as a second system
with equal temperature but activity $z_2$ has $N_2$ particles and $M_2$ contacts
is simply the product $P=P_{N_1 M_1}(\tau,z_1) P_{N_2 M_2}(\tau,z_2)$.  The
equilibrium probability of observing a pair of systems under the same conditions
but with the configurations exchanged is $P'=P_{N_2 M_2}(\tau,z_1) P_{N_1 M_1}(\tau,z_2)$.
Since the probability of attempting a configuration exchange is independent of
configuration, the correct joint distribution is sampled if exchanges are
accepted with probability
\begin{equation}
p^{\rm acc}_{\rm ex}={\rm min}[1,P'/P]={\rm min}[1,(z_1/z_2)^{N_2-N_1}].
\label{pex}
\end{equation}
The activities must be chosen close enough that typical values of
 $p^{\rm acc}_{\rm ex}$ are not too small.  As the extensivity of the exponent in
Eq.~(\ref{pex}) makes clear, $p^{\rm acc}_{\rm ex}$ decreases with increasing
system size because of the decreasing relative size of the fluctuations.
It is therefore necessary to use a larger number of parallel runs with more
closely spaced activities for larger systems.
\par
A natural companion of the parallel tempering technique is multiple
histogram reweighting,\cite{Ferrenberg88a,Ferrenberg89a} which allows
the results from simulations of different conditions to be combined.  One
may then obtain rigorously interpolated data for conditions different
from those explicitly simulated without the need for further runs.
Our implementation of this method is detailed in the Appendix.

\subsection{Inherent Pathology\label{pathology}}
It has been pointed out that for systems of $N\ge12$ adhesive hard
spheres, the configuration integral $Z_N$ diverges,\cite{Stell91a,Borstnik97a}
and that the model is therefore pathological.  The problem arises from
certain maximally connected clusters in which the number of contacts between
particles exceeds the number of vibrational degrees of freedom, leading to
non-integrable singularities in $Z_N$.
\par
The present simulations are concerned with the fluid phases of Baxter's model.
Under conditions where dense clusters form, aggregation tends to proceed in a
polytetrahedral manner, with particles attaching to existing triangular facets
due to the contact adhesion.  The clusters that lead to the divergence of
$Z_N$ are close-packed and characteristic of the crystal phase.  In contrast,
regular tetrahedra are not space-filling, and so troublesome clusters are
increasingly unlikely to arise spontaneously in the simulation as $\tau$
decreases.  Indeed, for such a short-range potential, the liquid phase is
expected to be highly metastable with respect to the crystal,\cite{Hagen93a}
with a high free energy barrier between the two.  Here
we effectively neglect the diverging contributions of crystal nuclei
containing 12 or more particles.
\par
In the event that a pathological cluster were to arise in the simulation,
the algorithm described in this section would simply fail to observe the
additional contact between particles that leads to non-integrability.
Since such ``accidental'' contacts arise from the geometrical packing of
strictly monodisperse spheres, the algorithm effectively introduces an
infinitesimal polydispersity.  For implications regarding the
applicability of the model, we refer readers to the detailed discussion
by Stell.\cite{Stell91a}

\section{Results\label{results}}

\subsection{Coexistence Curve}

A grand canonical simulation yields a histogram of the number
of particles in the system---effectively the probability distribution
of the reduced particle density, $\rho=\sigma^3 N/L^3$, where
$L$ is the length of the cubic simulation box.  We work throughout
with $\rho$, which is related to the volume fraction $\eta$ of the
spheres by $\eta=\rho\pi/6$.
\par
At any subcritical temperature, there is an activity at which the
probability distribution of $\rho$ is bimodal with peaks of equal
area, signifying equilibrium between a low- and a high-density fluid
phase.  The coexisting densities are given by the mean of each peak.
In practice, the finite size of the simulation box means that for
modestly subcritical temperatures, the two peaks overlap, and it
becomes impossible to attribute the intermediate densities to one
peak or the other.  It is therefore more reliable to define equilibrium
by equal heights of the peaks, and the coexisting densities by the
peak positions.
\par
We note that, at significantly subcritical temperatures,
where two separated peaks are observed, the effect of small changes
in the activity about the coexistence point is to alter the relative
heights of the peaks without affecting much their positions.  Under these
conditions, the positions of the peaks at coexistence are also not very
sensitive to the system size.  Closer to the critical point, discrepancies
due to the definition of coexistence and finite size effects both
increase.  However, using the peak heights is far less ambiguous than using
their areas.
\par
Table \ref{binodal} reports the coexisting (peak) densities as a function
of the temperature parameter $\tau$ in a cubic simulation box of length
$L=8\sigma$.  Also given are an estimate of the statistical error and a
measure of the size dependence of the results by comparison with simulations
at $L=6\sigma$.  At the lowest temperatures, where the coexisting density
peaks are well separated, the difference between the peak positions and their
integrated means is roughly 0.01, the direction of the differences
being such that the means of the two peaks are closer than their peak positions.

\begin{table}
\caption{\label{binodal}
Coexisting peak densities $\rho_{\rm lo}$ and $\rho_{\rm hi}$ of the low- and
high-density fluid phases of the adhesive hard sphere model with simulation cell
length $L=8\sigma$.  $\delta_{\rm ran}$
denotes the uncertainty in locating the peak positions.  $\delta_{\rm sys}$
is the average of the decrease in $\rho_{\rm lo}$ and the increase in
$\rho_{\rm hi}$ when the simulation cell size is {\em decreased} to $L=6\sigma$.
This quantity is an indication of the increasing size-dependence of the results
as the critical point is approached from below.}
\begin{ruledtabular}
\begin{tabular}{lllll}
$\tau$ & $\rho_{\rm lo}$ & $\rho_{\rm hi}$ & $\delta_{\rm ran}$ & $\delta_{\rm sys}$\\
\hline
  0.090 & 0.076 &  0.860 &  0.01  &  0.005 \\
  0.092 & 0.084 &  0.857 &  0.01  &  0.005 \\
  0.094 & 0.092 &  0.848 &  0.01  &  0.005 \\
  0.096 & 0.102 &  0.842 &  0.01  &  0.005 \\
  0.098 & 0.113 &  0.840 &  0.01  &  0.005 \\
  0.100 & 0.125 &  0.830 &  0.01  &  0.005 \\
  0.102 & 0.137 &  0.826 &  0.005 &  0.005 \\
  0.104 & 0.152 &  0.818 &  0.005 &  0.01  \\
  0.106 & 0.170 &  0.807 &  0.005 &  0.015 \\
  0.108 & 0.192 &  0.795 &  0.005 &  0.02  \\
  0.110 & 0.221 &  0.783 &  0.01  &  0.03  \\
  0.112 & 0.256 &  0.752 &  0.01  &  0.03  \\
  0.113 & 0.279 &  0.736 &  0.01  &  0.04  \\
  0.114 & 0.31  &  0.72  &  0.03  &  0.06  \\
  0.115 & 0.34  &  0.67  &  0.04  &  0.1   \\
\end{tabular}
\end{ruledtabular}
\end{table}

\begin{figure}
\includegraphics[width=85mm]{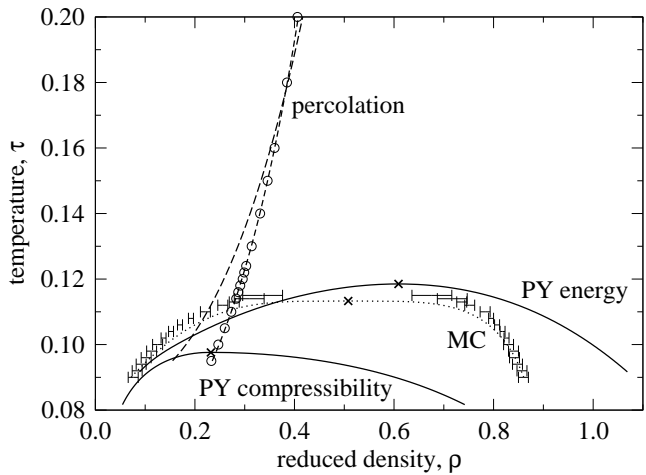}
\caption{\label{phasediagram}
Phase diagram of the adhesive hard sphere fluid.  The two solid lines are the
fluid--fluid coexistence curves from Percus--Yevick theory as marked.  The
simulation coexistence data from Table \ref{binodal} are shown by points with
error bars (marking statistical uncertainty),
with a dotted line merely to guide the eye.  Critical points are
indicated by crosses.  The dashed lines show the percolation threshold:
long dashes are the Percus--Yevick result, Eq.~(\ref{PYperc}) and short dashes
with circles are the simulation results and their fit, Eq.~(\ref{percfit}).
To the high-density side of the threshold the system essentially always contains
an infinite cluster.
}
\end{figure}

The pronounced size dependence near the top of the low- and high-density
branches was noted in previous work,\cite{Miller03a} and was exploited to
derive the critical point in the thermodynamic limit, as listed in the last line of
Table \ref{critical}.  At this point, the adhesive hard sphere density
distribution can be scaled onto the well-known\cite{Tsypin00a} universal
critical order-parameter distribution of the three-dimensional Ising model
at any system size\cite{Bruce92a,Wilding92a,Wilding95a}
($5\le L/\sigma\le 10$ were studied).
However, it is clear from Table \ref{binodal} and Fig.~\ref{phasediagram}
that the finite-system coexistence branches significantly overshoot the critical point.
It is partly because of the varying finite-size effects that
we have refrained from fitting the coexistence curve to a functional form,
such as a scaling law with universal exponents.
We also point out that the parameter $\tau$ is not
the thermodynamic temperature in the usual sense.  Even if
temperature is the relevant parameter in an experimental realization
of adhesive hard spheres, its relation to $\tau$ depends entirely on
the substance, and could be linear,\cite{Piazza98a} inverse,\cite{Mallamace00a}
or more complicated.\cite{Chen94a}

\begin{table}
\caption{\label{critical}
Critical temperature and density of the adhesive hard sphere model from the
Percus--Yevick compressibility\cite{Baxter68a} and energy\cite{Watts71a} routes,
and from simulation.\cite{Miller03a}
}
\begin{ruledtabular}
\begin{tabular}{lll}
method & $\tau_c$ & $\rho_c$ \\
\hline
PY compressibility & 0.0976 & 0.232 \\
PY energy & 0.1185 & 0.609 \\
simulation & $0.1133\pm0.0005$ & $0.508\pm0.01$
\end{tabular}
\end{ruledtabular}
\end{table}

Table \ref{critical} and Fig.~\ref{phasediagram} emphasize the large discrepancies in
the critical point and coexistence curve derived from the compressibility\cite{Baxter68a}
and energy\cite{Watts71a} routes of Percus--Yevick theory; the critical densities
differ by almost a factor of three.  The present simulations indicate that the true
properties lie between the two theoretical results, but somewhat closer to the energy
route, despite the fact that the compressibility results are more often used
when modeling experimental data.

\subsection{Percolation Threshold}

In principle, a percolating cluster is a dynamical object.
A system-spanning cluster may assemble, but its constituent
particles are in dynamic equilibrium with the rest of the system.
In a simulation, one can therefore define a percolation
probability equal to the fraction of the encountered
configurations that contain a system-spanning cluster. At a given
temperature, this probability is a sigmoidal function of density,
rising from zero to unity in a relatively narrow range.  The
density at which the probability passes through 0.5 has been found
to be quite insensitive to the system
size,\cite{Kranendonk88a,Lee01c} and we therefore adopt this point
as a robust definition of the percolation density, $\rho_{\rm
perc}(\tau)$.  For $\rho>\rho_{\rm perc}$ the system effectively
always contains a percolating cluster.
\par
We have measured $\rho_{\rm perc}(\tau)$ in canonical simulations of a system with
$N=500$ particles.  The decreasing density at which percolation sets in as $\tau$
is lowered means that the ergodicity problems associated with dense low-temperature
regime do not arise,
and parallel tempering is not necessary.  Our data are well reproduced by a ratio
of polynomials with six adjustable parameters:
\begin{equation}
\rho_{\rm perc}(\tau)=\frac{-10.09+182.4\tau+606.9\tau^2+15.31\tau^3}{1+507.9\tau+548.9\tau^2},
\label{percfit}
\end{equation}
applicable in the range $0.095\le\tau\le3$.  The maximum residuals between the measured
and fitted densities are 0.001 for the portion above $\tau=0.2$ and 0.002 below it.  These
errors are equal to the estimated uncertainty in the simulation data, as ascertained by
independent simulations with different random number seeds.
\par
Finite-size effects are relatively small in the percolation curve.  Reducing the system
size to $N=256$ particles shifts the curve systematically to higher density by about 0.001
towards the high-$\tau$ end of the range studied and 0.003 at the low-$\tau$ end.
\par
\begin{figure}
\includegraphics[width=85mm]{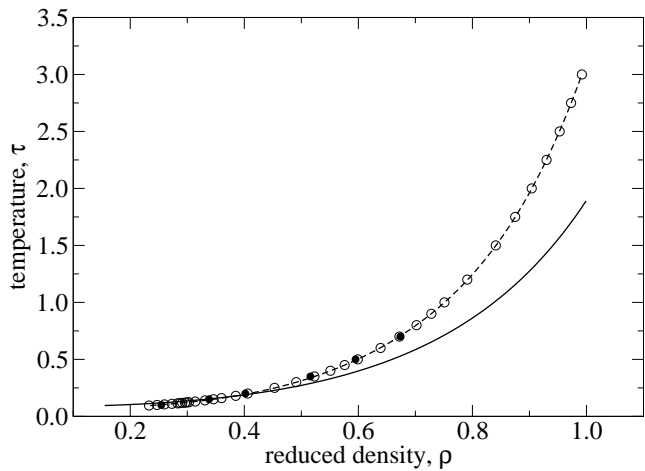}
\caption{\label{threshold}
Percolation threshold of the adhesive hard sphere fluid.  Open circles are Monte Carlo
data from this work, and the dashed line is the fit Eq.~(\ref{percfit}).  Filled circles
are Monte Carlo data from Table I of Ref.~\onlinecite{Lee01c}.  The solid line is the
Percus--Yevick result.}
\end{figure}

Figure \ref{threshold} shows the simulation data and the fit in Eq.~(\ref{percfit}) over
the full range studied.  We note that, at high $\tau$, the threshold density enters
the two-phase crystal--fluid region of ordinary hard spheres ($0.939<\rho<1.038$).\cite{Frenkel02b}
The onset of hard sphere behavior places an upper limit on the value of $\tau$ to which
we can trace the percolation threshold.  No meaning should be attached to the empirical
fit in Eq.~(\ref{percfit}) outside the specified range.
\par
Lee\cite{Lee01c} has performed a detailed scaling analysis of percolation in adhesive
hard spheres in the range $0.1\le\tau\le0.7$, and Fig.~\ref{threshold} shows that the latter
results are consistent with the present work.  Also plotted is the Percus--Yevick percolation
threshold derived by Chiew and Glandt from the divergence of the mean cluster size with
temperature at a given density:\cite{Chiew83a}
\begin{equation}
\tau_{\rm perc}(\eta)=\frac{19\eta^2-2\eta+1}{12(1-\eta)^2},\quad \eta=\rho\pi/6.
\label{PYperc}
\end{equation}
This formula increasingly overestimates the percolation density at given temperature for
$\tau>0.177$, but underestimates it at lower temperatures approaching the two-phase
regime (Fig.~\ref{phasediagram}).

Importantly, Fig.~(\ref{phasediagram}) shows that the critical point for the fluid--fluid
phase transition lies well within the percolated regime, in contrast to the results
of the Percus--Yevick compressibility results
that suggest the critical point is rather close to the
threshold.  In an experimental system with very short range
attractive forces, the onset of infinite clusters might therefore
interfere with equilibrium phase separation and critical behavior.\cite{Poon97a}

\subsection{Equation of State}

In the grand canonical ensemble, the pressure can be derived from the average density
$\langle\rho(z,\tau)\rangle$ as a function of the activity $z$ and temperature $\tau$.
For a change in chemical potential at constant volume and temperature,
$VdP=\langle N\rangle d\mu$.  Integration yields an isotherm of the reduced pressure
$P^*=\sigma^3P/kT$ as a function of the reduced density:
\begin{equation}
P^*(\rho,\tau)=\int_{z=0} \langle\rho(z,\tau)\rangle d(\ln z).
\label{isotherm}
\end{equation}
Below the critical point, one must integrate the mean densities of the two fluid phases
separately, determining the constant of integration for the high-density branch by
the requirement that the pressures be equal at the coexistence activity.
\par
The multiple histogram technique is invaluable in evaluating Eq.~(\ref{isotherm}),
since it allows the density to be found effectively as a continuous function of
the activity from a series of simulations at discrete points.  Equally, one can
interpolate between simulations at two values of $\tau$ to obtain an intermediate
isotherm.
\par
We have performed grand canonical parallel tempering simulations spanning a wide range
of particle densities ($0<\rho<1$) at each of $\tau=0.1$, 0.105, 0.11, 0.12, 0.13,
0.15, 0.2, 0.25, 0.35, 0.5, 1, 1.5, 2, 3, and 5 using a simulation box of length
$L=8\sigma$.  These directly-measured isotherms were supplemented by 22 intermediate
temperatures derived by histogram reweighting from the pair of simulations bracketing
each temperature.
\par
The rather unconventional definition of the adhesive hard sphere model in
Eq.~(\ref{potential}) has consequences for fitting of an empirical equation of
state to the data $P(\rho,\tau)$.  Firstly, the high temperature limit is
not an ideal gas but a system of hard spheres, whose equation of state is reproduced
accurately by Speedy's formula:\cite{Speedy97a}
\begin{equation}
P^*_{\rm HS} = \rho\left[1+\frac{x+0.076014x^2+0.019480x^3}{1-0.548986x+0.075647x^2}\right],
\label{speedy}
\end{equation}
where $x=B_2^{\rm HS}\rho$ and $B_2^{\rm HS}=2\pi/3$ is the reduced second virial
coefficient of hard spheres.
Additionally, the first four virial coefficients of adhesive hard spheres are known
analytically,\cite{Post86a} consisting of the corresponding temperature-independent
hard-sphere part plus a contribution from the adhesion.  The adhesive contributions are
\begin{eqnarray}
B_2^{\rm adh}(\tau) &=& -\frac{B_2^{\rm HS}}{4\tau}, \label{b2adh}\\
B_3^{\rm adh}(\tau) &=& \left(\frac{B_2^{\rm HS}}{4}\right)^2\left(-5\tau^{-1}+\tau^{-2}-\frac{1}{18}\tau^{-3}\right),
\label{b3adh}\\
B_4^{\rm adh}(\tau) &=& \left(\frac{B_2^{\rm HS}}{4}\right)^3 (-13.77358\tau^{-1} +6.114\tau^{-2}\nonumber\\
   && -1.518\tau^{-3} + 0.17398\tau^{-4} \nonumber\\
   && -6.33514\times10^{-3}\tau^{-5} -6.51271\times10^{-5}\tau^{-6}).\label{b4adh}
\end{eqnarray}
With this information, it is already clear that many existing empirical
equations of state are not of an appropriate form for this model.  Even the popular
modified Benedict--Webb--Rubin (mBWR) equation,\cite{Nicolas79a}
which employs 33 adjustable parameters and has been successfully fitted
to the Lennard-Jones fluid equation of state,\cite{Johnson93a} is
unsatisfactory as it stands: the terms containing fractional powers of
temperature are redundant in the light of Eqs.~(\ref{b2adh}) and (\ref{b3adh}),
and its high temperature limit is not flexible enough to reproduce Eq.~(\ref{speedy}).
\par
We have attempted to adapt the mBWR equation by removing unnecessary terms
and inserting ones more appropriate to the adhesive hard sphere model.  However,
we have been unable to obtain a satisfactory fit across the full gamut of our
simulation data and must therefore be satisfied with an empirical fit to
the supercritical part of the equation of state, which we now present.  We note that
the supercritical regime is relevant to many experimental applications of the
adhesive hard sphere model.
\par
To guarantee the correct low-density (up to third order in $\rho$) and high-temperature
limits, we write the full adhesive hard sphere pressure as
\begin{equation}
P^*(\rho,\tau) = P^*_{\rm HS}(\rho) + B_2^{\rm adh}(\tau)\rho^2 + B_3^{\rm adh}(\tau)\rho^3 +
P^*_{\rm fit}(\rho,\tau),
\label{psplit}
\end{equation}
and fit the final term to a ratio of mixed polynomials in $\rho$ and $1/\tau$:
\begin{equation}
P^*_{\rm fit}(\rho,\tau)=\frac{\sum_{\{a_{ij}\}}a_{ij}\rho^i/\tau^j}
{1+\sum_{\{b_{ij}\}}b_{ij}\rho^i/\tau^j}.
\label{eofsfit}
\end{equation}
In Eq.~(\ref{eofsfit}) the sums are over the sets of terms, with coefficients
$\{a_{ij}\}$ and $\{b_{ij}\}$, that have been chosen to provide the flexibility to
reproduce the data.
\par
We have used a standard nonlinear least-squares routine\cite{Press92a} to optimize
Eq.~(\ref{eofsfit}) to reproduce our simulated and interpolated isotherms $P^*(\rho,\tau)$
for the $L=8\sigma$ system
at 35 values of $\tau\ge0.117$ (the apparent finite-system critical point
lying just below $\tau=0.117$).  Starting from a modest number of terms, contributions
were inserted and removed according to their ability to improve the fit, and the final
result was a set of 35 terms, 24 in the numerator and 11 in the denominator, as listed
in Tables \ref{numerator} and \ref{denominator}.  We emphasize that the fit
is meaningless below the lowest temperature used in the parametrization, $\tau=0.117$,
but may be applied at arbitrarily high $\tau$ since Eq.~(\ref{psplit}) reverts
smoothly to the ordinary hard sphere equation of state at $\tau=\infty$.

\begin{table}
\caption{\label{numerator}
Numerator coefficients for the equation of state, Eq.~(\ref{eofsfit}).
}
\begin{ruledtabular}
\begin{tabular}{lld@{\qquad}lld}
$i$ & $j$ & a_{ij} & $i$ & $j$ & a_{ij}\\
\hline
4 & 2 & -3.787821    & 6 & 4 & -0.1954175 \\
4 & 4 & -0.3184060   & 6 & 5 & -0.3832949 \\
4 & 5 & 0.1460575    & 7 & 1 & -207.9268 \\
4 & 6 & -0.01199764  & 7 & 2 & 301.5938 \\
5 & 1 & -24.12376    & 7 & 3 & -88.02139 \\
5 & 2 & 0.6172855    & 7 & 4 & 6.609781 \\
5 & 4 & -1.679166    & 8 & 1 & 114.6900 \\
5 & 5 & 0.05973859   & 8 & 2 & -155.9904 \\
5 & 6 & 0.01606935   & 8 & 3 & 42.26076 \\
6 & 1 & 111.7043     & 8 & 4 & -4.117703 \\
6 & 2 & -137.3363    & 8 & 5 & 0.1512075 \\
6 & 3 & 43.98811     & 8 & 6 & -0.003129547 \\
\end{tabular}
\end{ruledtabular}
\end{table}

\begin{table}
\caption{\label{denominator}
Denominator coefficients for the equation of state, Eq.~(\ref{eofsfit}).
}
\begin{ruledtabular}
\begin{tabular}{lld@{\qquad}lld}
$i$ & $j$ & b_{ij} & $i$ & $j$ & b_{ij}\\
\hline
1 & 2 & 29.25140   & 3 & 2 & 85.86019 \\
1 & 3 & -3.259845  & 3 & 3 & -8.978804 \\
2 & 1 & 32.11475   & 4 & 1 & 24.33100 \\
2 & 2 & -88.09174  & 4 & 2 & -26.90679 \\
2 & 3 & 9.484818   & 4 & 3 & 2.751372 \\
3 & 1 & -57.07652 \\
\end{tabular}
\end{ruledtabular}
\end{table}

Despite the inapplicability of the fit below $\tau=0.117$, one can always obtain the
pressure in the low density limit using the
hard-sphere equation of state plus the exact adhesive contributions up to fourth
order in $\rho$ as listed in Eqs.~(\ref{b2adh})--(\ref{b4adh}):
\begin{equation}
P^*(\rho,\tau) = P^*_{\rm HS}(\rho) + B_2^{\rm adh}(\tau)\rho^2 + B_3^{\rm adh}(\tau)\rho^3 +
B_4^{\rm adh}(\tau)\rho^4.
\label{lowrho}
\end{equation}
In fact, the low-density side of the coexistence curve is sufficiently rarefied that
Eq.~(\ref{lowrho}) is quite satisfactory for the full gas-like branch of
subcritical isotherms, except for $\tau$ very close to the critical point.
\par
\begin{figure}
\includegraphics[width=85mm]{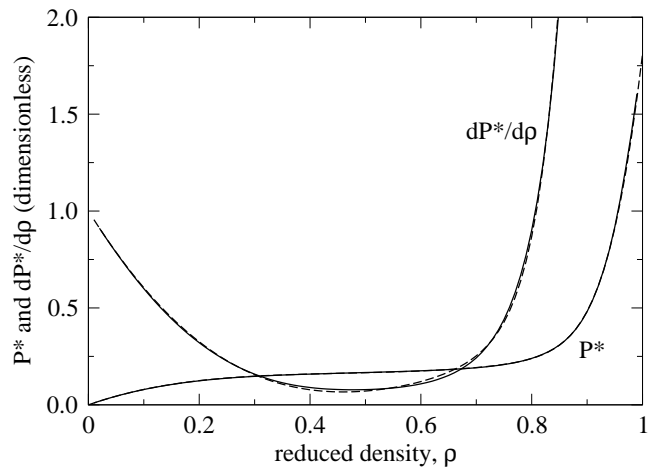}
\caption{\label{tau012}
Comparison of the $\tau=0.12$ isotherm from simulation and as
reproduced by Eqs.~(\ref{psplit}) and (\ref{eofsfit}).
Also shown is the dimensionless derivative $(\partial P^*/\partial \rho)_\tau$
from both simulation and
fit.  In each case, the solid line is the simulation data and the dashed
line the fit, though the two are not everywhere discernible.
}
\end{figure}

Returning to the supercritical equation of state, Fig.~\ref{tau012} shows the pressure
along the $\tau=0.12$ isotherm to demonstrate
the performance of the fit.  Note that the simulation results can be shown
as a continuous line, since the multiple histogram technique allows us to calculate
the pressure on an arbitrarily fine grid.  The figure also shows the (dimensionless)
derivative $(\partial P^*/\partial \rho)_\tau$, which is related to the dimensionless
isothermal compressibility $\kappa^*$ by $1/\kappa^*=\rho(\partial P^*/\partial\rho)_\tau$.
The fit reproduces the derivative acceptably, despite the derivative not being used in
the fitting process.  In fact, Fig.~\ref{tau012} is a somewhat pessimistic case, since
the agreement between the simulation data and fit, especially for the
derivative, improves with increasing $\tau$.

\begin{figure}
\includegraphics[width=85mm]{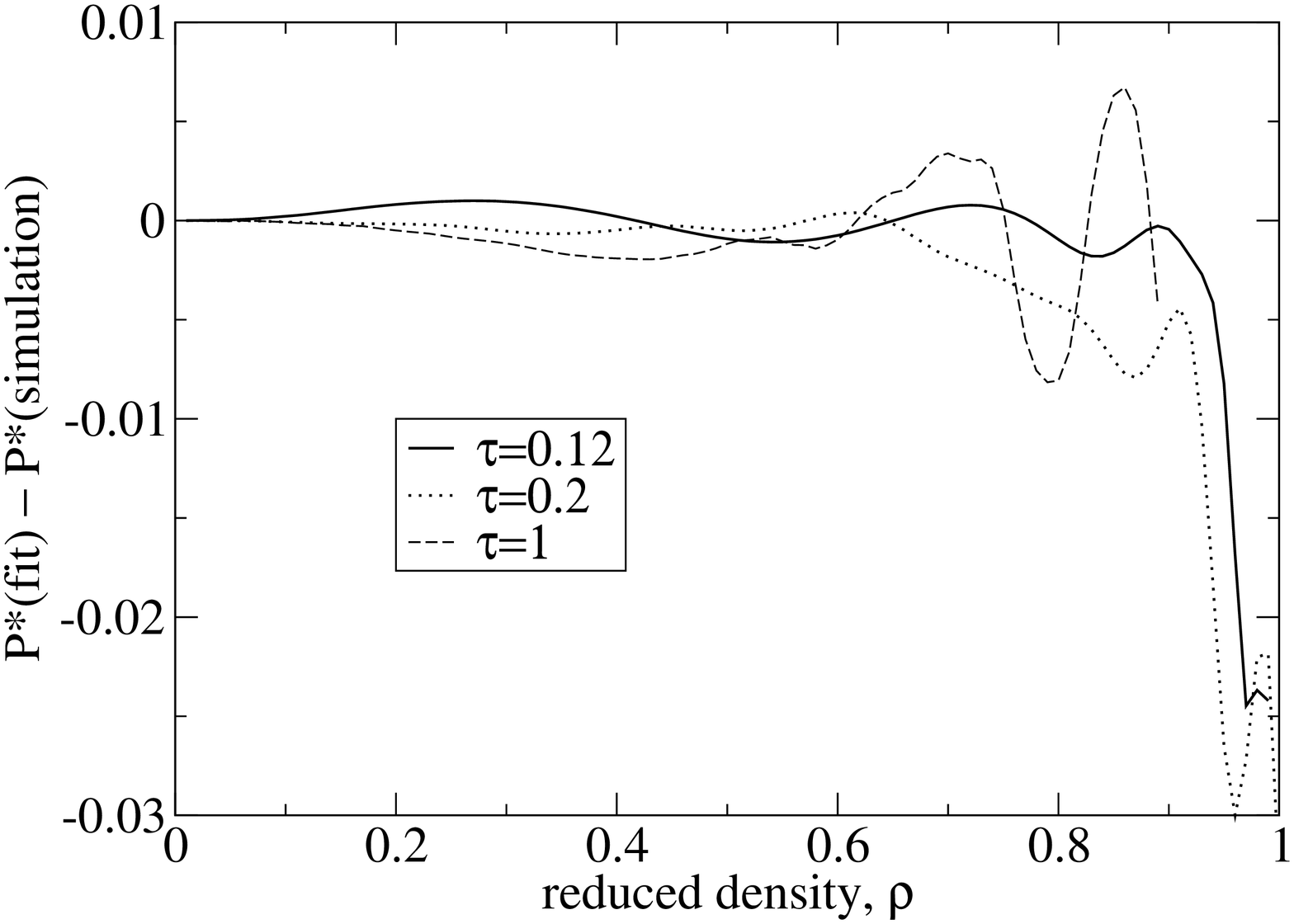}
\caption{\label{errors}
Discrepancy (unscaled differences) between the pressure directly from
simulations and the empirical form Eq.~(\ref{eofsfit}) for three temperatures.
}
\end{figure}

The actual (not fractional) difference between the simulation and the fit is
plotted for three temperatures in Fig.~\ref{errors}.  By construction, the
discrepancy goes smoothly to zero at zero density.  The increasing absolute
size of the discrepancy at high densities still represents a small fractional
error in the light of the steeply rising pressure in that regime (see Fig.~\ref{tau012}).
Approaching the critical point, the simulation data and fit suffer from
finite-size effects.  However, simulations with a smaller simulation cell of $L=6\sigma$
produce almost indistinguishable results for $\tau\ge0.14$, so the equation
of state provided here is robust over a wide range of temperatures.

\begin{figure}
\includegraphics[width=85mm]{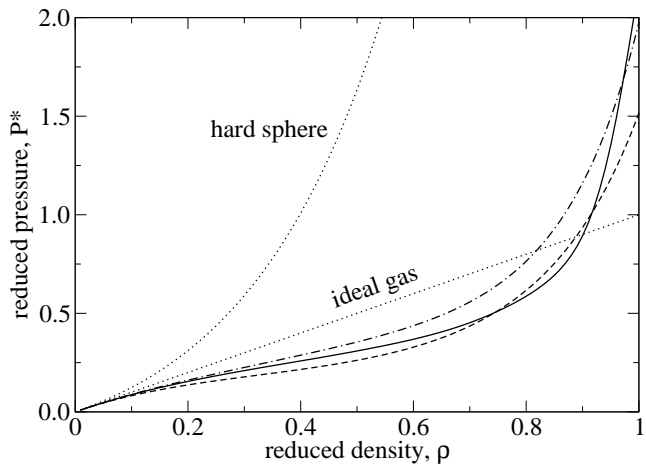}
\caption{\label{supcritisotherm}
Adhesive hard sphere pressure on a supercritical isotherm.
The solid curve is the simulation result at $\tau=0.15$, equivalent to
$\tau/\tau_c=1.3239$.  The dashed and dot-dashed lines are the Percus--Yevick
compressibility and energy results, respectively, each calculated at
the same $\tau/\tau_c$ using the associated $\tau_c$.  Shown by
dotted lines are the ideal gas and ordinary hard sphere equations of state.
}
\end{figure}

We now compare the simulation isotherms with the predictions of Percus--Yevick
theory.  Expressions for the pressure derived from the compressibility and
energy routes, originally due to Baxter\cite{Baxter68a} and Watts et
al.,\cite{Watts71a} are given in convenient and explicit form by
Barboy\cite{Barboy74a} and Tenne.\cite{Barboy79a}  Figure \ref{supcritisotherm}
shows the pressure along a supercritical isotherm at $\tau/\tau_c=1.3239$,
which corresponds to $\tau=0.15$ using the critical temperature from
simulations.  The two Percus--Yevick results are plotted for the same value of
the ratio $\tau/\tau_c$, but using their respective critical temperatures
(Table \ref{critical}).  Also plotted for comparison are the ideal gas
equation of state and Eq.~(\ref{speedy}) for ordinary hard spheres.  We
see that the attractive component of the adhesive hard sphere potential can
dramatically reduce the pressure with respect to hard spheres even at high density.
Indeed, the attraction at this temperature is strong enough to overpower the
effects of excluded volume to the extent that the adhesive hard sphere pressure
lies below that of the ideal gas up to $\rho\approx0.9$ (simulation data).

\begin{figure}
\includegraphics[width=85mm]{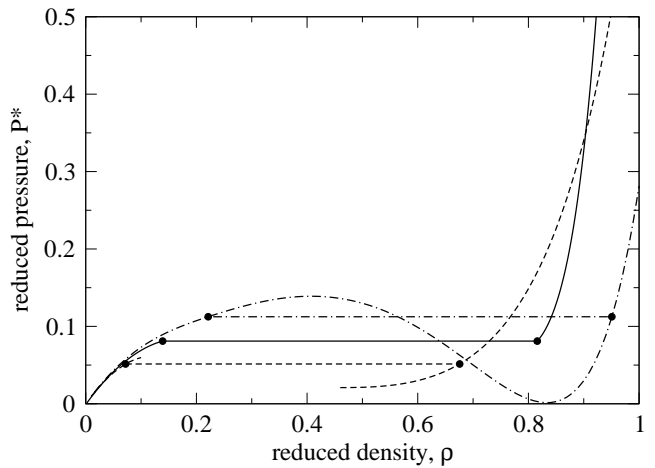}
\caption{\label{subcritisotherm}
Adhesive hard sphere pressure on a subcritical isotherm.
The solid curve is the simulation result at $\tau=0.1$, equivalent to
$\tau/\tau_c=0.8826$.  The dashed and dot-dashed lines are the Percus--Yevick
compressibility and energy results, respectively, each measured at the
same $\tau/\tau_c$ using the associated $\tau_c$.  Tie lines join coexisting
densities.
}
\end{figure}

Figure \ref{subcritisotherm} compares subcritical isotherms at $\tau/\tau_c=0.8826$,
corresponding to $\tau=0.1$ for the simulation data.  Coexisting densities have been
joined with tie lines.  There is a gap in the compressibility equation curve where the
expression for the pressure has no real solution.  Scaling the temperatures according
to the associated critical points places the simulation results for the low-density
branch pressure and the coexisting densities between the two routes of Percus--Yevick
theory.  We note, however, that the Percus--Yevick pressure is rather unrealistic in
the high-density branch, especially from the energy equation, since the pressure must
rise very steeply at sufficiently high density.

\section{Concluding Remarks}

It is noteworthy that, despite there being at least two analytical equations
of state for the adhesive hard sphere model, arising from the compressibility
and energy routes of Percus--Yevick theory, it is by far more common for the
compressibility results to be cited and used.  It is therefore significant
that, at least as far as the phase
coexistence curve is concerned, the present results place the energy
equation closer to the truth.
\par
In summary, we have presented the first firm numerical results for the
fluid--fluid coexistence curve of the adhesive hard sphere model, as well as
accurate empirical fits to the percolation threshold and the supercritical
equation of state.
Given the popularity of Baxter's model in the interpretation of experimental
data in colloid science, we hope that this information
will enhance the model's utility.  The results also constitute a reference for
future simulations of short ranged attractive systems.

\begin{acknowledgments}
The authors gratefully acknowledge insightful discussions and correspondence
with Prof.~W.~C.~K.~Poon and Prof.~G.~Stell.
The work of the FOM Institute is part of the research program of
FOM and is made possible by financial support from the Netherlands
organization for Scientific Research (NWO).
\end{acknowledgments}

\appendix

\section{Multiple Histogram}

Recall Eq.~(\ref{pnm}) for the probability of observing a system of adhesive
hard spheres with $N$ particles and $M$ contacts under specified conditions
of effective temperature $\tau$ and activity $z$.
The continuous probability distribution of the density, $P(\rho; \tau, z)$
can be approximated from the discrete
probability in Eq.~(\ref{pnm}) by
summing over $M$ and normalizing with respect to integration over $\rho=\sigma^3 N/V$:
\begin{equation}
P(\rho;\tau,z)=\frac{\sigma^3}{V}\sum_M P_{NM}(\tau,z).
\label{msum}
\end{equation}
A full knowledge of the density of states
$\Omega_{NM}$ to within a multiplicative constant
would therefore permit calculation of the density distribution at any
combination of temperature and activity using Eqs.~(\ref{pnm}) and
(\ref{msum}).
\par
In this appendix, we detail our implementation of the multiple
histogram technique\cite{Ferrenberg88a,Ferrenberg89a} for adhesive hard
spheres.  The aim is to combine
the data from several simulations at different temperatures and
activities to produce an optimal estimate for the density of
states over a wide range of $N$ and $M$ so that Eqs.~(\ref{pnm})
and (\ref{msum}) can be applied.  We follow closely the
approach described by Weerasinghe and Amar\cite{Weerasinghe93a}
for the canonical ensemble.  Our specific adaptation for the
adhesive hard sphere model and the grand canonical ensemble is included
here for completeness.
\par
Consider a grand canonical simulation at temperature $\tau_i$ and
activity $z_i$, where the subscript $i$ labels the simulation.
One can accumulate a two-dimensional histogram of
$N$ and $M$.  Let $S_{iNM}$ be the number of sampled configurations
that had $N$ particles and $M$ contacts, and let $s_i=\sum_{N,M} S_{iNM}$
be the total number of samples.  Abbreviating $\Xi(\tau_i,z_i)$ to
$\Xi_i$, the histogram furnishes the estimate
\begin{equation}
S_{iNM}/s_i \approx P_{NM}(\tau_i,z_i) = \Xi_i^{-1}\Omega_{NM}z_i^N\tau_i^{-M},
\label{hist}
\end{equation}
where the equality is approximate because of the statistical uncertainty inherent
in the simulation.
If $n_{\rm bins}$ combinations of $N$ and $M$ are sampled, the corresponding
$n_{\rm bins}$ values of $\Omega_{NM}/\Xi_i$ can be obtained from Eq.~(\ref{hist})
by straightforward division.  The points lying in the tails of the distribution
suffer the largest uncertainty due to poor sampling.
\par
If $n_{\rm sim}$ simulations are performed at different conditions
$(\tau_i,z_i)$, $1\le i\le n_{\rm sim}$, producing $n_{\rm sim}$ overlapping
two-dimensional histograms, we obtain $n_{\rm bins}\times n_{\rm sim}$
instances of Eq.~(\ref{hist}).  The task of the multiple histogram
technique is therefore to find the $n_{\rm sim}$
values of $\Xi_i$ and the $n_{\rm bins}$ values of $\Omega_{NM}$ that
minimize the overall discrepancy between the probabilities they define through
Eq.~(\ref{pnm}) and those measured in the various simulations.  By treating
these $n_{\rm sim}+n_{\rm bins}$ unknowns as individually adjustable
parameters, we will obtain the density of states and partition function
without making any assumptions about their functional forms.
\par
We start by taking the logarithm of Eq.~(\ref{hist}):
\begin{equation}
\ln S_{iNM} - \ln s_i \approx \ln\Omega_{NM} + N\ln z_i -M\ln\tau_i - \ln\Xi_i.
\label{logs}
\end{equation}
The left- and right-hand sides of Eq.~(\ref{logs}) are the measured and fitted
values of $\ln P_{NM}$, respectively.  The discrepancy is the residual,
\begin{equation}
R_{iNM}=\ln p_{iNM} - (\ln\Omega_{NM} + X_{iNM} -\ln\Xi_i),
\label{rinm}
\end{equation}
where we have introduced the abbreviations $p_{iNM}=S_{iNM}/s_i$ and
$X_{iNM} = N\ln z_i - M\ln\tau_i$.
\par
If we assume that sampled configurations are uncorrelated, the variance of $S_{iNM}$
measured in a simulation of finite length is expected to be\cite{Ferrenberg89a}
$\sigma^2(S_{iNM})=\langle S_{iNM}\rangle$, where the angle brackets indicate the
true ensemble average over simulations of the same length.
If we approximate this average by the value actually measured,
then we can estimate the standard deviation of $S_{iNM}$ to be
$\sigma(S_{iNM})=S_{iNM}^{1/2}$.  Propagating errors, we find
$\sigma(\ln p_{iNM})=S_{iNM}^{-1/2}$.  Summing the squares of the residuals divided
by their standard deviations, we arrive at the maximum likelihood estimator
\begin{equation}
\chi^2=\sum_i\sum_{N,M}S_{iNM}R^2_{iNM}.
\label{chisq}
\end{equation}
The first sum in Eq.~(\ref{chisq}) runs over the $n_{\rm sim}$ simulations, and
the second runs over all bins in the two-dimensional histogram.
\par
We now treat the $n_{\rm sim}$ values of $\Xi_i$ and the $n_{\rm bins}$ values of $\Omega_{NM}$
as adjustable parameters to be fitted by minimizing $\chi^2$.
Differentiating Eq.~(\ref{chisq}) with respect to the fitting variables and equating to
zero yields
\begin{alignat}{2}
\partial\chi^2/\partial\ln\Omega_{NM}&=-2\sum_i S_{iNM} R_{iNM}&=0,
\label{deriv1}\\
\partial\chi^2/\partial\ln\Xi_i&=2\sum_{N,M} S_{iNM} R_{iNM}&=0.
\label{deriv2}
\end{alignat}
\par
Substituting Eq.~(\ref{rinm}) into Eq.~(\ref{deriv1}) and solving for $\ln\Omega_{NM}$,
we obtain $n_{\rm bins}$ equations of the form
\begin{equation}
\ln\Omega_{NM} = \sum_i q_{iNM}(\ln p_{iNM} -X_{iNM} +\ln\Xi_i),
\label{sol1}
\end{equation}
where $q_{iNM}=S_{iNM}/\sum_j S_{jNM}$.  Similarly, substituting Eq.~(\ref{rinm})
into Eq.~(\ref{deriv2}) and solving for $\ln\Xi_i$, we obtain $n_{\rm sim}$
equations of the form
\begin{equation}
\ln\Xi_i = \sum_{N,M} p_{iNM}(\ln\Omega_{NM}-\ln p_{iNM} + X_{iNM}).
\label{sol2}
\end{equation}
\par
We now have expressions for the partition functions exclusively in terms of the
density of states and vice versa.  Since there are fewer of the former than the
latter, we solve for the partition functions first.  Substituting
Eq.~(\ref{sol1}) into Eq.~(\ref{sol2}) and yields
\begin{multline}
\ln\Xi_i=\sum_{N,M}p_{iNM}\bigg[-\ln p_{iNM} + X_{iNM}\\
+\sum_j q_{jNM} (\ln p_{jNM} -X_{jNM} +\ln\Xi_j)\bigg].
\end{multline}
Gathering terms in $\Xi_i$ on the left-hand side gives
\begin{multline}
\ln\Xi_i-\sum_{N,M}p_{iNM}\sum_j q_{jNM}\ln\Xi_j \\
=\sum_{N,M}p_{iNM}\sum_j(q_{jNM}-\delta_{ij})(\ln p_{jNM}-X_{jNM}),
\label{finalsub}
\end{multline}
where $\delta_{ij}$ is the Kronecker delta.
\par
Eq.~(\ref{finalsub}) is a set of $n_{\rm sim}$ linear equations in $n_{\rm sim}$
unknowns, and can be written in matrix notation as
\begin{equation}
{\bf AY}={\bf B},
\label{matrix}
\end{equation}
where $Y_i=\ln\Xi_i$.  The elements of the matrix ${\bf A}$ are
\begin{equation}
A_{ij}=\delta_{ij}-\sum_{N,M}p_{iNM}q_{jNM},
\label{matrixa}
\end{equation}
and the elements of the vector ${\bf B}$ are
\begin{equation}
B_i=\sum_{N,M}p_{iNM}\sum_j(q_{jNM}-\delta_{ij})(\ln p_{jNM}-X_{jNM}).
\label{vectorb}
\end{equation}
\par
One practical point in evaluating ${\bf A}$ and ${\bf B}$ is that
many of the $S_{iNM}$ are zero, making $\ln p_{iNM}$ and $\ln q_{iNM}$
incalculable.  However, such logarithms always appear in products involving
$S_{iNM}$ itself.  Since $\lim_{x\to0} x\ln x=0$, bins with zero contents
simply make no contribution to Eqs.~(\ref{chisq}), (\ref{matrixa}) and
(\ref{vectorb}).
\par
Solution of Eq.~(\ref{matrix}) leads to numerical values for the partition
functions, which can then be substituted into Eq.~(\ref{sol1}) to obtain the
density of states.  In principle, however, one cannot obtain the absolute
values of the partition functions from a set of equations like (\ref{pnm}),
only their ratios.  For this reason, the matrix ${\bf A}$ is singular, and
Eq.~(\ref{matrix}) should be tackled using singular-value
decomposition,\cite{Press92a} yielding $\Xi_i$ and $\Omega_{NM}$ up to
a multiplicative constant.

\end{document}